\documentclass{aa} 
\usepackage[varg]{txfonts}
\usepackage{graphicx}

\usepackage{color,soul} 

\newcommand\un[1]{{\,\rm #1}}
\newcommand\E[1]{\times10^{#1}}
\newcommand\rs[1]{_\mathrm{#1}}
\newcommand\pd[2]{\frac{\partial{#1}}{\partial{#2}}}
\newcommand\g{$\gamma$}

\begin{document} 

\title{Linking gamma-ray spectra of supernova remnants to the cosmic ray injection properties in the aftermath of supernovae}
\titlerunning{Time-dependent injection and \g-rays}

\author{O. Petruk \inst{1,2} \and S. Orlando\inst{1} \and M. Miceli\inst{3,1} \and F.Bocchino\inst{1}}

\institute{
INAF - Osservatorio Astronomico, Piazza del Parlamento, 1, 90134 Palermo, Italy
\and
Institute for Applied Problems in Mechanics and Mathematics, Naukova Street, 3-b, 79060 Lviv, Ukraine
\and 
Dipartimento di Fisica e Chimica, Universit\'a degli Studi di Palermo,  Viale delle Scienze, 17, 90128 Palermo, Italy
}

\date{Received ...; accepted ...}

\abstract{
The acceleration times of the highest-energy particles which emit gamma-rays in young and middle-age SNRs are comparable with SNR age. If the number of particles starting acceleration was varying during early times after the supernova explosion then this variation should be reflected in the shape of the gamma-ray spectrum. We use the solution of the non-stationary equation for particle acceleration in order to analyze this effect. As a test case, we apply our method to describe gamma-rays from IC443. As a proxy of the IC443 parent supernova we consider SN1987A. First, we infer the time dependence of injection efficiency from evolution of the radio spectral index in SN1987A. Then, we use the inferred injection behavior to fit the gamma-ray spectrum of IC443. We show that the break in the proton spectrum needed to explain the gamma-ray emission is a natural consequence of the early variation of the cosmic ray injection, and that the very-high energy gamma-rays originate from particles which began acceleration during the first months after the supernova explosion. We conclude that the shape of the gamma-ray spectrum observed today in SNRs critically depends on the time variation of the cosmic ray injection process in the immediate post explosion phases. With the same model, we estimate also the possibility in the future to detect gamma-rays from SN 1987A.
}


\maketitle

\section{Introduction}

There is a gap in observations which prevents us to see how the properties of supernovae (SNe) -- asymmetry of explosion, initial structures in distribution of ions, in circumstellar medium  etc. -- influence their further evolution toward the supernova remnants (SNRs) and how they determine SNR morphologies. There are just few known supernova events occurred in our Galaxy during the centuries: we cannot therefore track directly the development of such object. It merits effort therefore to look for other possible relationships between SNe and SNRs which are observed at the present time. One of such ideas is described here. 

Typically, the interpretation of the nonthermal emission from SNRs involves a solution of the steady-state equation for the particle acceleration. The time-dependent acceleration however could be important in analysis of young and middle-aged SNRs because the system eventually has not enough time to reach the steady-state regime. There are numerical simulations which demonstrate that the particle spectrum is not stationary even in the rather old SNRs \citep{Brose-et-al-2016}.
The solution of the time-dependent equation demonstrates that the variable injection (fraction of particles which begin acceleration) affects the slope of the power-law part and the shape of the high-energy end of the CR spectrum \citep{Petruk-Kopytko-2016}. 

Diffusive shock acceleration increases the particle energy on a small amount in each acceleration cycle. In order to reach the high energy, a particle should diffuse around the shock for a long time. 
Ultrarelativistic electrons lose energy via synchrotron radiation. There are evidences that in several SNRs their maximum energy is loss-limited. In particular, the loss-limited scenario has been successfully adopted to describe the X-ray spectrum of RX J1713.7-3946 \citep{Zirak-Aha-2010,Tanaka-etal-2008}, of Tycho \citep{Morlino-Caprioli-2012} and of the nonthermal limbs of SN 1006 \citep{Miceli-etal-2013}. At odds with electrons, hadrons do not suffer significant radiative losses and the highest energy particles are mostly those which have the maximum available time for acceleration. They have to start accelerating very early, around or shortly after the time of supernova explosion. Therefore, the time variation of the number of particles which enter the acceleration during the first decades after the supernova explosion affects the high-energy end of the particle spectrum and thus the hadronic \g-ray emission of SNRs. 

In the present paper, we adopt the measurements of the time variation of the radio spectral index in a SN in order to reconstruct the time evolution of the particle injection efficiency during the early times after the explosion. The extracted dependence is then used in order to derive the particle momentum distribution and the \g-ray spectrum of a SNR. 
Here we challenge our method considering the radio observations of SN1987A and the \g-ray spectrum of SNR IC443.

\section{Method}
\label{alpha:sect-equation}

\subsection{Time-dependent solution}

The distribution of particles in space and momentum with ongoing diffusive shock acceleration is described by the isotropic non-stationary distribution function $f(t,x,p)$. 
The equation for its evolution is \citep{Skilling1975a,Jones-1990}:
\begin{equation}
 \pd{f}{t}+u\pd{f}{x}=\pd{}{x}\left[D\pd{f}{x}\right]+\frac{1}{3}\frac{du}{dx}p\pd{f}{p}+Q
 \label{kineq:kineq}
\end{equation}
where $t$ is the time, $x$ the spatial coordinate, $p$ the momentum, $D$ the diffusion coefficient, $u$ the flow velocity in the shock reference frame, $Q$ the injection term;  
the velocities of the scattering centers are neglected. 
We assume injection to happen at the shock, to be isotropic and monoenergetic with the initial momentum $p\rs{i}$:
\begin{equation}
 Q(t,x,p)=\frac{\eta n_1u_1}{4\pi p\rs{i}^2}\delta(p-p\rs{i})\delta(x)Q\rs{t}(t),
 \label{kineq:umova2b}
\end{equation}
where the parameter $\eta$ is the injection efficiency (the density fraction of accelerated particles); actually, it is the amplitude of the efficiency at the saturated level. 
The term $Q\rs{t}(t)$ represents the time evolution of the injection (that may be caused by the variation of $\eta$ or $n_1$). It is unity in the steady-state regime. 
In the present paper, the indices '1', '2' and 'o' correspond to upstream, downstream and to the shock itself. 

The test-particle solution of equation (\ref{kineq:kineq}) for the steady-state injection $Q\rs{t}=1$ was derived by \citet{Drury-1983,Forman-Drury-1983}; the authors additionally assume that the upstream acceleration time $t_1=4D_1/u_1^2$ is much higher than the downstream one $t_2=4D_2/u_2^2$. More general solution for the time-dependent injection and any relation between $t_1$ and $t_2$ was derived by \citet{Petruk-Kopytko-2016}. It was shown that if $t_1/t_2$ is larger than few, the difference with the solution obtained for the limit $t_1\gg t_2$ is negligible. Therefore,  for simplicity, we consider $t_1\gg t_2$ in the present paper. 
In that case, the particle distribution function at the shock which satisfies the equation (\ref{kineq:kineq}) and accounts for the variable injection $Q\rs{t}(t)$ is \citep{Petruk-Kopytko-2016}
\begin{equation}
 f\rs{o}(t,p)=\mathrm{f}\rs{o}(p)\int\limits_{0}^{\tau} Q\rs{t}(\tau-\tau') \varphi\rs{o}(\tau') d\tau'.
\label{kineq2:solfTPQ}
\end{equation}
In this expression, 
\begin{equation}
 \mathrm{f}\rs{o}(p)=\frac{\eta n_1}{4\pi p\rs{i}^3}\frac{3\sigma}{\sigma-1}\left(\frac{p}{p\rs{i}}\right)^{-s\rs{f}},
 \label{kineq2:stationarysol}
\end{equation}
is the solution of the stationary equation, 
\begin{equation}
 s\rs{f}={3\sigma}/{(\sigma-1)},
\end{equation}
is the spectral index, $\sigma=u_1/u_2$ the shock compression factor and \citep{Forman-Drury-1983,Petruk-Kopytko-2016}
\begin{equation} 
 \varphi_{\mathrm{o}}(\tau)=\frac{e^{2A}}{2^{2A+1}\sqrt{\pi}}\frac{e^{-\xi(\tau)^2}}{\tau^{A/2+1}}
 \left(\mathrm{H}_{A+1}\left(\xi\right)
 -{2\tau^{1/2}}\mathrm{H}_{A}\left(\xi\right)
 \right),
 \label{kineq2:t1phi}
\end{equation}
where $\mathrm{H}_m(x)$ is the Hermite polynomial, $\tau=t/t_1$, $\xi(\tau)=\tau^{1/2}+A/(2\tau^{1/2})$, $A=s\rs{f}/\alpha$, $\alpha$ the index in the dependence of the diffusion coefficient on the particle momentum $D(p)\propto p^{\alpha}$ (the value of $\alpha$ should provide $A$ to be integer). 

At this point we can answer the following questions: Is and to which extend the time-dependent consideration of the particle acceleration are really necessary? Is the steady-state acceleration regime reasonable to describe the radio and $\gamma$-ray emission? 
The equation (\ref{kineq2:solfTPQ}) clearly shows that the time-dependent solution equals the steady-state one if the integral over $\tau$ is unity. 
This may happen starting from some $\tau$, if the injection is constant, i.e. $Q\rs{t}=1$. 
Thus, the criterion when we may adopt the steady-state description of the particle acceleration is
\begin{equation}
 {\cal I}(\tau)\equiv \int_{0}^{\tau} \varphi\rs{o}(\tau') d\tau' \approx 1. 
\end{equation}

\begin{figure}
 \centering
 \includegraphics[width=7truecm]{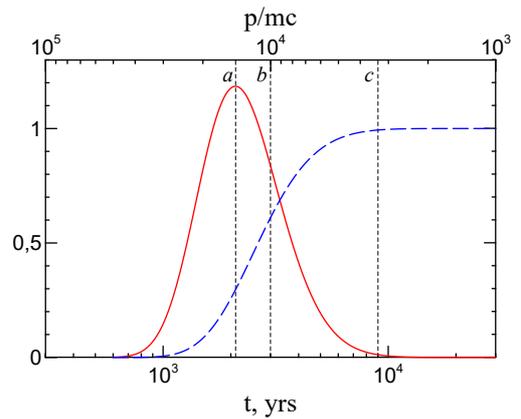}
 \caption{Probability $\varphi\rs{o}$ (red solid line) and the integral ${\cal I}$ (blue dashed line) versus time for the fixed momentum $p=p\rs{max}$ (bottom horizontal axis) and versus particle momenta for the fixed time $t=t\rs{age}$ (top horizontal axis). Numbers on the horizontal axes correspond to the particular choice of $t\rs{age}=3000\un{yrs}$,  $p\rs{max}=10^{4}m\rs{p}c$ and $\alpha=1$. Meaning of the dotted vertical lines is explained in the text.
               }
 \label{alpha:figprob}
\end{figure}

The function $\varphi\rs{o}(\tau(t,p))$ is the two-dimensional probability distribution for particles injected with the momentum $p\rs{i}$ at the time $t\rs{i}$ to be accelerated to the momentum $p$ in time $t$. Let us consider, without loss of generality of the conclusions, an SNR with age $t\rs{age}=3000\un{yrs}$ and the maximum particle momentum $p\rs{max}=10^{4}m\rs{p}c$ and plot two cross sections of the 2D distributions $\varphi\rs{o}$ and ${\cal I}$: for the fixed momentum $p=p\rs{max}$ and for the fixed time $t=t\rs{age}$  (Fig.~\ref{alpha:figprob}). Looking at the top axis we see that: i) during the time $t\rs{age}$, most of the particles (maximum of $\varphi\rs{o}$) are accelerated to the momentum (marked by the dotted vertical line {\it a}) somehow larger than $p\rs{max}$ (which is marked by the dotted vertical line {\it b}); ii) particles with momenta smaller than the threshold momentum $\approx p\rs{max}/3$ (marked by the vertical line {\it c}) may be described by the steady-state solution. The bottom axis on Fig.~\ref{alpha:figprob} demonstrates that: i) most particles are accelerated to the momentum $p\rs{max}$ during the time (vertical line {\it a}) which is somehow smaller than $t\rs{age}$ (vertical line {\it b}); ii) the acceleration regime for particles around the maximum momentum is not yet steady-state because ${\cal I}(p\rs{max})=0.6$ for the time $t\rs{age}$; iii) the acceleration of particles with momentum $p=p\rs{max}$ may be considered to be steady-state after the threshold time (vertical line {\it c}) which is about 3 times larger than $t\rs{age}$. 

In summary, the time-scale for the steady acceleration till $p\rs{max}$ is few times larger than the SNR age. Therefore, if injection is constant, it is mandatory to consider the time-dependent acceleration for particles with the highest momenta; and these particles are actually those emitting $\gamma$-rays in SNRs. If injection varies in time, the time-dependent regime has to be adopted for all particles which start acceleration before the injection reaches the level $Q\rs{t}=1$.

\subsection{Our approach}

The solution (\ref{kineq2:solfTPQ}) allows us to consider effects of the variable injection on the spectrum of accelerated particles and on their nonthermal emission. 

Our approach is the following. We use properties of some supernova observed in the present epoch to reconstruct the early evolution of some supernova remnant, assuming that they have similar evolution during the first years after SN event. 

Namely, we start from the observations of the time variation of the radio spectral index $\alpha\rs{r}$ in the SN. 
This index is known to be $\alpha\rs{r}=(s-3)/2$ where $s$ is the spectral index of the particle distribution $f\rs{o}$. 
Thus, at the first step, we find the function $Q\rs{t}(\tau)$ by solving the integro-differential equation 
\begin{equation}
 \alpha\rs{r}(t)=-\frac{3}{2}-\left.\frac{1}{2}\frac{d\ln f\rs{o}(t,p)}{d\ln p}\right|_{p=p\rs{*}},
 \label{alpha:aradiodef}
\end{equation}
where $f\rs{o}(t,p)$ is given by (\ref{kineq2:solfTPQ}), $\alpha\rs{r}(t)$ are taken from observations and $p\rs{*}$ is the particle momentum which gives the maximum contribution to emission at the observational frequency. 
As the result, we uncover how the injection efficiency evolved during the early times after the SN explosion. 

At the second step, we use this function $Q\rs{t}(\tau)$ as an input to the solution (\ref{kineq2:solfTPQ}), in order to model the particle spectrum $f\rs{o}(t\rs{*},p)$ in the SNR at the present time $t\rs{*}$. Having the particle spectrum, we calculate the nonthermal emission of the SNR, in particular, its \g-ray spectrum. 

There is a property noted by \citet{Petruk-Kopytko-2016} which essentially simplifies the first step. 
Namely, if the injection term is of the form $Q\rs{t}\propto \tau^\beta$, then 
the spectral index of the radio emitting electrons in SNRs is approximately 
\begin{equation}
 s=s\rs{f}+\alpha\beta
 \label{alpha:sradiosnr}
\end{equation}
(for details see Appendix \ref{alpha:app1}).
We use this relation to reconstruct $Q\rs{t}(t)$ without solving the equation (\ref{alpha:aradiodef}). 
In fact, we derive the time series of different $\beta_i$ from the observed series of $\alpha_{\mathrm{r}i}$ and approximate $Q\rs{t}(t)$ by a complex piecewise function. 
The normalization of this function comes from the condition that $Q\rs{t}=1$ in the steady-state part.  

The solution (\ref{kineq2:solfTPQ}) is expressed through the unitless time $\tau$ which, in fact, is a function of $t$ and $p$. Therefore, we need two scaling rules, in order to convert between the physical and normalized values. The first rule is obviously  
\begin{equation}
 t=\tau t\rs{m}, \quad t\rs{m}=t_1(p\rs{*}).
 \label{alpha:tm}
\end{equation}  
The second rule is given by the definition of $\tau$ written for a general $p$ in the form
\begin{equation}
 \tau=\frac{tu_1^2}{4D_1(p\rs{max})}\frac{p\rs{max}^\alpha}{p^\alpha};
\end{equation} 
it follows from here that
\begin{equation}
 p=\tau^{-1/\alpha}p\rs{m},\quad p\rs{m}=\frac{t\rs{age}^{1/\alpha}p\rs{max}}{t_1(p\rs{max})^{1/\alpha}}
 \label{alpha:pm}
\end{equation} 
where $t\rs{age}$ is the SNR age and $p\rs{max}$ is the maximum momentum of accelerated particles in the SNR which spectrum we would like to explain.  
The value of $t_1(p\rs{max})$ is a rough estimate of the acceleration time to reach the momentum $p\rs{max}$. Since the time available is limited by the age of SNR then $t_1(p\rs{max})\simeq t\rs{age}$ and therefore $p\rs{m}\simeq p\rs{max}$.

In the present paper, we set $\sigma=4$ that corresponds to the unmodified shock and $\alpha=1$ which is relevant to the Bohm-like diffusion. 

\begin{figure}
 \centering
 \includegraphics[width=8.4truecm]{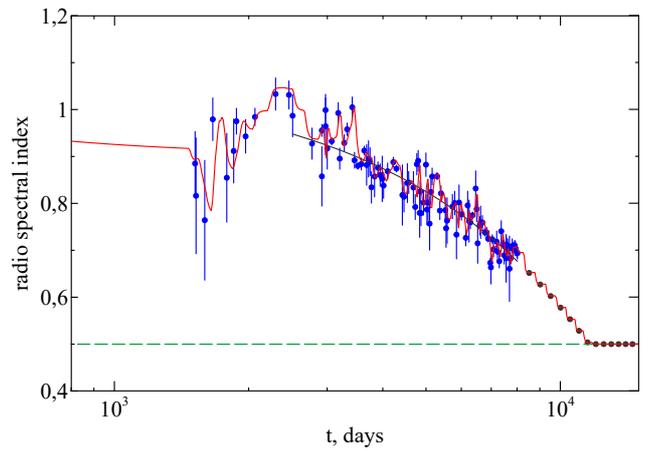}
 \caption{Evolution of the radio spectral index as observed by \citet{Zanardo-etal-2010} during 1517 - 8014 days (filled circles with errors), extrapolation (black circles without errors) of the linear fit (black line; from the same reference) up to the value $\alpha\rs{r}=0.5$; then the constant $\alpha\rs{r}$ is assumed. Red solid line represents the spectral index $\alpha\rs{r}$ calculated by Eq.~(\ref{alpha:aradiodef}) with the dependence $Q\rs{t}(t)$ derived from the observed $\alpha\rs{r}$ with the simple approximate formula (\ref{alpha:sradiosnr}); this $Q\rs{t}(t)$ is shown on Fig.~\ref{alpha:figq}. Green dashed line represents the case of the steady-state injection. 
               }
 \label{alpha:figalpha}
\end{figure}

\section{Ongoing acceleration in SN1987A and the \g-ray spectrum of IC443}
\label{alpha:sect-sn1987a}

The described approach is adopted in the present paper to the pair of the supernova 1987A and the supernova remnant IC443.  
SN1987A is a type II supernova \citep{Lymanetal-2014}. IC443 is generally expected to be also from the type II SN. The initial arguments for this were based on SNR location in the star-forming region and the association with a neutron star and its pulsar-wind nebula \citep{Bocchino-Bykov-2001,Olbert-etal-2001,Gaensler-etal-2006,Swartz-etal-2015}. The low Fe and high Mg abundance \citep{Troja-etal-2008} as well as the position of the centroid of the Fe K$\alpha$ emission \citep{Yamaguchi-etal-2014} found by observing X-rays from IC443 strengthen this view. As discussed below, our results also support the core-collapse scenario for this SNR. 

\subsection{Radio index of SN1987A}
\label{alpha:sectradio}

Evolution of the radio spectral index in SN1987A is presented by \citet{Zanardo-etal-2010}, from ATCA observations during 1517 to 8014 days after explosion. We plot it on Fig.~\ref{alpha:figalpha} (filled circles with errors). The spectral index does not reach the saturated value (something close to the canonical $0.5$) yet, thus the injection in SN1987A is not in the steady state. In order to trace a possible future evolution of the spectral index, we extrapolate the linear fit of the data up to the time when the spectral index become 0.5 and then keep it constant (black circles without errors), as shown on Fig.~\ref{alpha:figalpha}.

\begin{figure}
 \centering
 \includegraphics[width=8.2truecm]{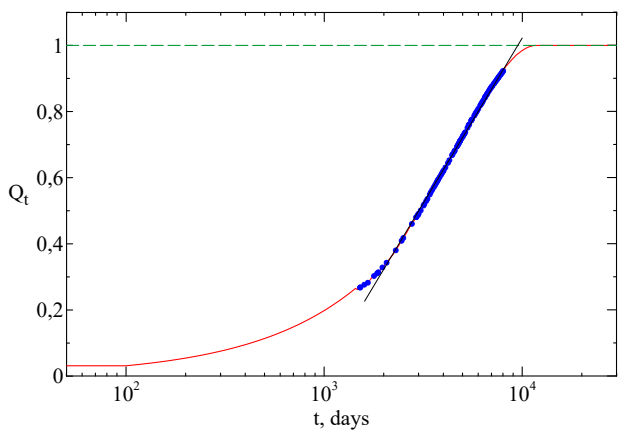}
 \caption{Time dependence of injection term. Blue circles correspond to the days of the radio observations. Red solid line traces the function $Q\rs{t}(t)$ derived as described in Sect.~\ref{alpha:sect-equation} from the observations during 1517 - 8014 days and from the extrapolation of $\alpha\rs{r}$ after the day 8014. As to the behavior of the injection term before the day 1517, we adopt a simple approximation as described in the text, with $\beta\rs{x}=0.8$ and $d\rs{x}=100$.
 Green dashed line represents the case of the steady-state injection. Thin black line is a function $Q\rs{t}=\lg(t/950)$. 
               }
 \label{alpha:figq}
\end{figure}

At the first step, the time-dependence of the injection term is derived from $\alpha\rs{r}$ with the use of the approximate formula (\ref{alpha:sradiosnr}), as described in the previous section. It is shown on Fig.~\ref{alpha:figq}. It is interesting that the derived function $Q\rs{t}(t)$ looks smooth despite of the noisy behavior of the radio index. In particular, during most of the observational period it may approximately be traced by the simple dependence $Q\rs{t}=\lg(t/(950\un{days}))$. This means that $Q\rs{t}$ is sensitive to the average global shape of the radio index evolution. All short-time features, that may depend on the details of the blast wave interaction with nonuniformities of the circumstellar medium, do not significantly affect the shape of $Q\rs{t}$.

SN1987A was invisible in radio before the day 1200 \citep{Zanardo-etal-2010}. Therefore, we describe the behavior of $Q\rs{t}(t)$ at earlier times with a simple approximation: it is steady-state from the beginning up to the day $d\rs{x}$ and then $Q\rs{t}(t)\propto t^{\beta\rs{x}}$ after this day to the day 1450, then it follows the dependence from observations. The values of the parameters $\beta\rs{x}$ and $d\rs{x}$ are determined ad hoc, from the requirement to provide acceptable fit of the \g-ray spectrum of IC443. It is worth to mention that there are detailed radio observations of SN1987A also during the first months after the explosion \citep[from February to September, e.g.][]{Ball-etal-2001} that are useful for the studies of the flux evolution. Papers which report the radio spectral index, which is of interest for our purposes, \citep{turtle-etal-1987,storey-manchester-1987} demonstrated that the radio spectrum at this early stage is affected by the free-free absorption and the synchrotron self-absorption that makes it impossible to use these data in the solution of the equation (\ref{kineq:kineq}) which does not account for these effects.

In order to check accuracy of our approach, we calculated the spectral index $\alpha\rs{r}$ from the full Eq.~(\ref{alpha:aradiodef}) using derived $Q\rs{t}$. One can see on Fig.~\ref{alpha:figalpha} (solid red line) that we restore even the small-scale features of the original observational data, so our approximate approach which bases on Eq.~(\ref{alpha:sradiosnr}) is quite accurate and robust.

\subsubsection{What could be a reason of the injection behavior?}
\label{alpha:reason}

Though it is not necessary for the purpose of the present paper -- which consists in derivation of the time-dependence of the injection from observations and in its use in modeling the gamma-ray emission -- we investigate the reason of the obtained time dependence of the injection efficiency.

\begin{figure}
 \centering
 \includegraphics[width=6.2truecm]{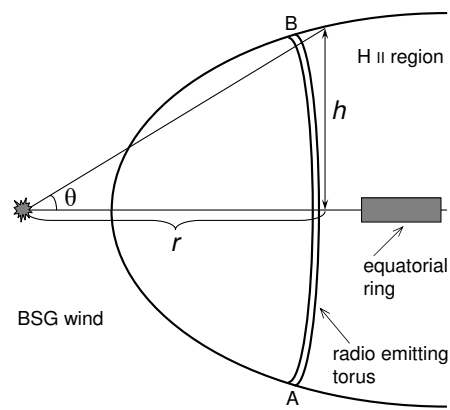}
 \caption{The structure of the ambient medium around SN1987A in the 3D numerical model from \citet{orlando-etal-2015}. The shock is almost plane-parallel between points A and B because it decelerates in the dense medium of the H~\textsc{ii} region.
               }
 \label{alpha:figsxema}
\end{figure}

\begin{figure*}
 \centering
 \includegraphics[width=16truecm]{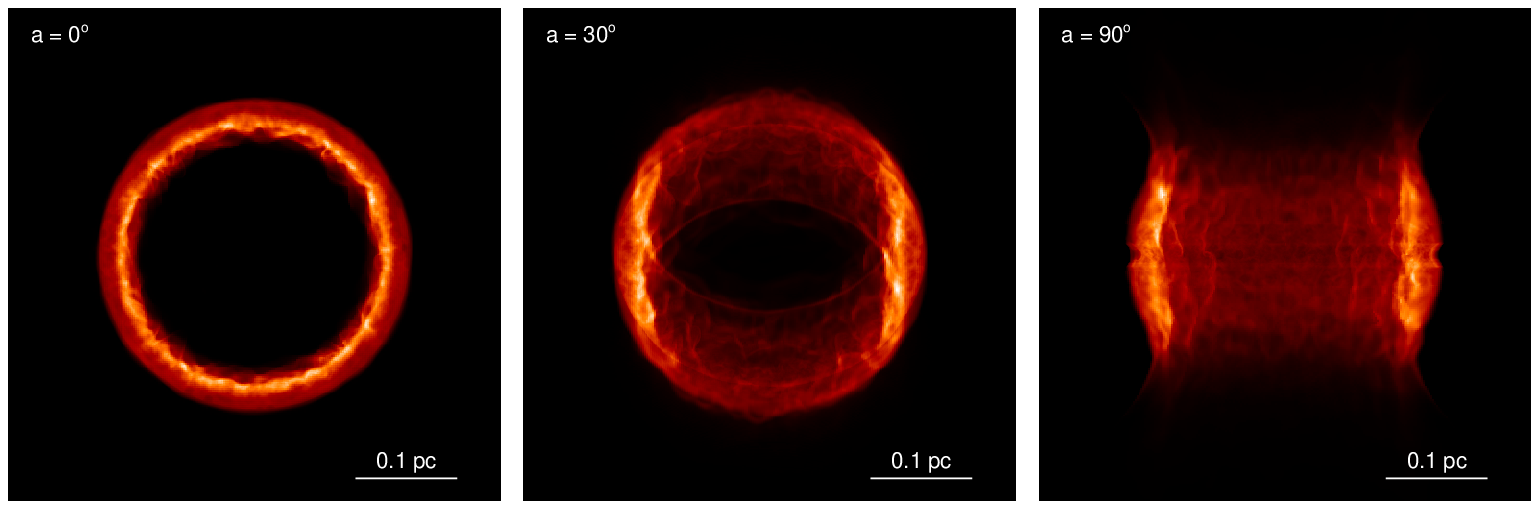}
 \caption{3D volumetric rendering of the density in the numerical model of SN1987A \citet{orlando-etal-2015}, at 15 years (around Day 5500) after the supernova explosion. The three panels show the rendering at three different angles (see upper left corner of each panel) between the line of sight and the axis of the H~\textsc{ii} region surrounding SN1987A. The colorscale is normalized to the maximum value in each panel.
               }
 \label{alpha:fig3D}
\end{figure*}

\begin{figure}
 \centering
 \includegraphics[width=8truecm]{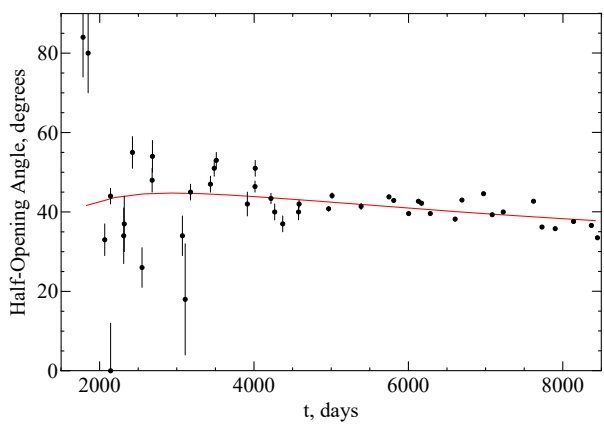}
 \caption{The temporal evolution of the half-opening angle $\theta$ (black dots with errors) derived by analysis of observations \citep{ng-etal-2013} and from the 3D numerical model \citep{orlando-etal-2015} (red solid line). 
               }
 \label{alpha:figtheta}
\end{figure}

It is clear from observations that SN1987A may not be approximated by an emitting sphere \citep[e.g.][]{ng-etal-2013,zanardo-etal-2013}. The accurate analysis of the radio observations in the Fourier space \citep{ng-etal-2008,ng-etal-2013} reveals that the radio emission arises from a ring-like structure. Detailed 3D numerical models of SN1987A \citep{potter-etal-2014,orlando-etal-2015} supported this fact and have demonstrated that this emitting torus is a consequence of the interaction of the supernova shock with the H~\textsc{ii} region with density few orders of magnitude higher than in the blue super-giant wind where the rest of the SN evolves. Fig.~\ref{alpha:figsxema} illustrates a sketch of the ambient medium structure and Fig.~\ref{alpha:fig3D} shows the 3D rendering of the post-shock density as derived from the model which explains the temporal and spatial properties of the X-ray emission from SN1987A \citep{orlando-etal-2015}.\footnote{At this point, our use of the equation~(\ref{kineq:kineq}), which describes the plane shock, is justified: we divide the thin emitting cylinder on a number of sectors and consider the shock to be plane-parallel in each sector.} In order to quantify the changes of the thickness of the emitting ring, \citet{ng-etal-2008,ng-etal-2013} introduced the half-opening angle $\theta$ (see our Fig.~\ref{alpha:figsxema}) and derived its temporal evolution (shown by the black dots with errors on Fig.~\ref{alpha:figtheta}). Actually, the thickness of the H~\textsc{ii} region in the 3D numerical model of \citet{orlando-etal-2015} fits well the evolution of $\theta$ (shown by the red solid line on Fig.~\ref{alpha:figtheta}). The thickness of the H~\textsc{ii} region is approximated by the function $h(r)= 0.25((r-r\rs{Hii})/r\rs{m})^{0.25}\un{pc}$ where $r$ is the distance from the explosion site in the equatorial plane, $r\rs{Hii}=0.08\un{pc}$ is the inner radius of the H~\textsc{ii} region, and the scale $r\rs{m}=3\un{pc}$.

\begin{figure}
 \centering
 \includegraphics[width=8truecm]{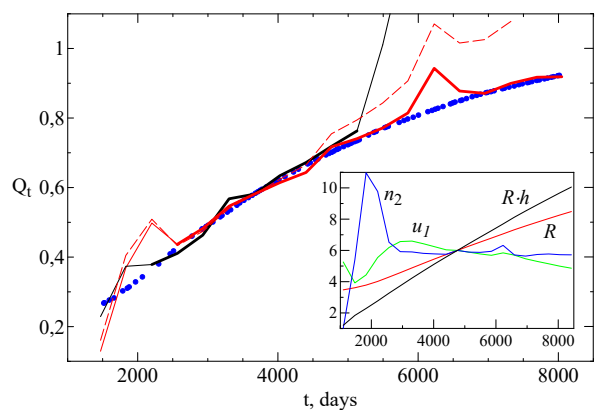}
 \caption{The temporal evolution of the injection efficiency derived from the radio spectral index (blue dots) and from the 3D model of SN1987A \citep{orlando-etal-2015}. 
The black (red) line represents $Q\rs{t,num}$ and corresponds to the case when the dense equatorial ring in the circumstellar nebula around SN1987A is accounted (neglected). The thick parts of the lines correspond to time when the shock propagates in the uniform H~\textsc{ii} region. The volume increase per unit time was calculated as $Rhu_1$ for the solid lines and as $R^2u_1$ for the dashed line. The rapid increase in the black line shortly after Day 5000 corresponds to the shock interaction with the dense equatorial ring in the nebula around the SN. The rapid increase and following decrease of the red line right after the Day 6000 is due to the few ejecta clumps which protrude the forward shock and causes an artificial increase of the post-shock density and velocity around this time. \textbf{Inset}: evolution of different characteristics, in arbitrary units. This plot corresponds to the red line on the main figure. One can see that the main contribution to the evolution of $Q\rs{t,num}$ comes from the surface increase, $R\cdot h$.
               }
 \label{alpha:figq2}
\end{figure}

From the same 3D numerical model, besides $h$, we can calculate the average characteristics of the shocked flow, namely, the shock radius $R$, the pre-shock flow velocity $u_1$ and the post-shock density of plasma $n_2$ which is proportional to the pre-shock density $n_1$. With these data, we may calculate the time evolution of the product
\begin{equation}
 Q\rs{t,num}(t) \propto n_1(t) u_1(t) R(t) h(t)
\label{alpha:Qtnum}
\end{equation} 
that is an approximate measure of the particle flux through the (increasing) shock surface. 

We considered two cases, namely, a uniform H~\textsc{ii} region with and without a denser equatorial ring inside. In order to mimic these two configurations with our single simulation, we have averaged the values in the cross-section in the equatorial plane (the black line on Fig.~\ref{alpha:figq2}) and in a plane inclined on few degrees (the red line on Fig.~\ref{alpha:figq2}) in order to simulate the situation when the shock does not interact with the dense equatorial ring in the nebula around SN1987A. Both the black and the red lines follow the 'observational' trend during the time when the shock propagates in the uniform H~\textsc{ii} region. Interestingly, it seems that the equatorial ring in the circumstellar medium does not contribute to the particle injection and to the radio emission (only the red line follows the blue dots after Day $5500$). This is in agreement with the results of \citet{potter-etal-2014}: blue and green lines on their Fig.~11, which represent the model accounting for the equatorial ring, does not follow the observed evolution of the radio flux after Day $5500$ when the shock encounters the equatorial ring. 

An important property is revealed by Fig.~\ref{alpha:figq2}: the 'observed' $Q\rs{t}$ (blue dots) follows the product $n_1u_1Rh$ derived from the numerical simulations (lines). This means that the fraction of particles injected into the Fermi acceleration process seems to be approximately the constant fraction of the particles entering the shock front. In turn, the flux of incoming particles changes during the radio observation period mostly (see inset on Fig.~\ref{alpha:figq2}) due to the shock surface increase. 

All SNRs expand very rapidly ('explosively') at the early times. In this sense the derived dependence $Q\rs{t}$ on time is 'general': it arises mainly from increasing surface of the blast wave (common for all SNRs) and does not depend significantly on individual
properties of SNe, at least during the first $\sim 30$ years of
evolution. In order to strengthen this conclusion even more, we
calculated the flux through the shock of SN 1987A either assuming
a cylindrical expansion (namely assuming the flux proportional to
$R h u_1$) of the remnant (red solid line in Fig.~\ref{alpha:figq2}) or a spherical
expansion (flux proportional to $R^2 u_1$; red dashed line). The
former assumption is motivated by the cylindrical geometry of the
blast wave evident from both observations and modeling of this SNR
(see Fig.~\ref{alpha:fig3D}). The second assumption is the most common for
the expansion of SN blast waves. Fig.~\ref{alpha:figq2} shows that the two cases
differ by no more than $\approx 20\%$ and both follow the same
general trend. Thus, we conclude that the particular geometry of
SN 1987A does not affect the generality of our analysis.

In fact, the increase of shock surface plays a prominent role in
the evolution of injection only during the early phases of SNRs and
becomes weaker at later times. The common scheme of SNR evolution
is that the shock moves freely after the explosion (with roughly
constant velocity), so that $u_1=\textrm{const}$, $R=u_1t$. Then
the shock decelerates in the later adiabatic stage and the Sedov
solutions give $u_1\propto t^{-3/2}$, $R\propto t^{2/5}$. As
a result, the dependence of the volume on time is: $R^2u_1\propto t^{2}$ 
during the free expansion, and $R^2u_1\propto t^{1/5}$ during
the adiabatic phase. In other words, the effects on the flux
increase is significantly reduced for SNRs which are not very young.

\begin{figure}
 \centering
 \includegraphics[width=8.6truecm]{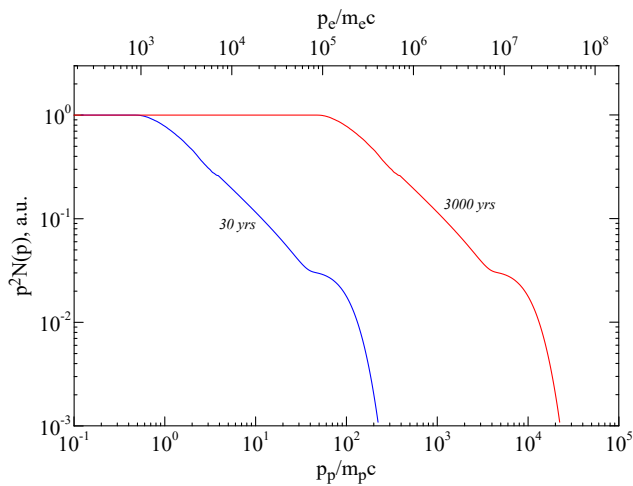}
 \caption{
Particle spectra at 30 and 3000 years calculated with Eq.(\ref{kineq2:solfTPQ}) and the dependence $Q\rs{t}(t)$ shown on Fig.~\ref{alpha:figq}. Bottom axis is for protons and the upper one is for electrons, under condition that both populations have the same maximum momentum (parameters used are listed in the caption to Fig.~\ref{alpha:figfpp}). 
               }
 \label{alpha:figsp}
\end{figure}

In our case, Fig.~\ref{alpha:figq} demonstrates that $Q\rs{t}$ reaches unity and is constant shortly after Day $10^4$ (i.e. sometime after $30$ years). In the case $Q\rs{t}=\textrm{const}$, our approach predicts that the injection does not affect  the spectrum slope of the particles incoming after this time and therefore their distribution function has the classic test-particle shape $p^{-s\rs{f}}$ (see Eqs.~\ref{kineq2:stationarysol} and \ref{alpha:sradiosnr}: constant $Q\rs{t}$ means $\beta=0$).
This is true for all momenta except the highest
ones, because the highest-energy particles are those injected at
the earliest times (they need more time for acceleration) when
injection (number of particles starting accelerations) was smaller. 
Fig.~\ref{alpha:figsp} illustrates how the particle spectra evolve with time.

Finally, it is worth to emphasize that our solution (\ref{kineq2:solfTPQ}) of the equation (\ref{kineq:kineq}) is obtained for the constant velocity $u_1$. However, comparing Eqs.~(\ref{kineq:umova2b}) and (\ref{alpha:Qtnum}), we see that the velocity variation is accounted in our solution through the term $Q\rs{t}$. In addition, the timescale of velocity change is small comparing to the particle acceleration rate. We expect therefore that our solution restores the main features in the particle spectrum while the time variation of velocity in other terms of the equation (\ref{kineq:kineq}) provide only the higher order corrections.

\subsection{\g-rays from IC443}

Once we have the evolution of injection in a supernova from the radio observations, we may proceed to the second step, namely, to use the derived $Q\rs{t}$ in order to calculate the particle spectrum and to simulate the \g-ray spectrum from IC443. In doing so, we assume that the injection of electrons and protons has the same time dependence (as on Fig.~\ref{alpha:figq}) and that the particle escape is negligible. 

\begin{figure}
 \centering
 \includegraphics[width=8.8truecm]{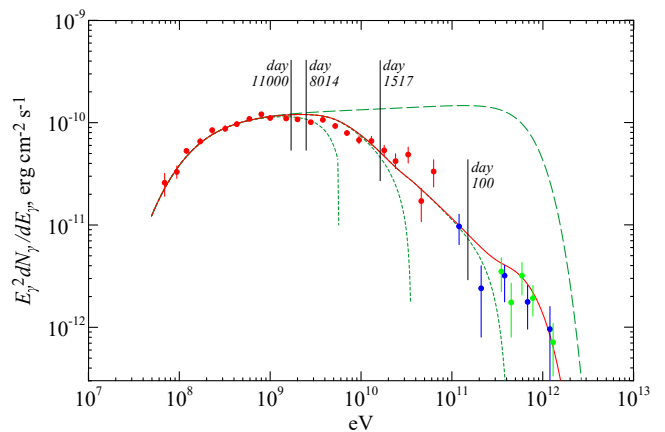}
 \caption{Gamma-ray spectrum of IC443.  
The FERMI data are shown by the red, MAGIC by blue and VERITAS by green dots with errors.    
Our fit is shown by the solid red line. The distribution of the parental protons is shown by the red line on Fig.~\ref{alpha:figsp}. The emission and the particle spectra are calculated with the scales $t\rs{m}=60\un{days}$ and $p\rs{m}=9.5\E{3}m\rs{p}c$ as well as the values of $d\rs{x}=100\un{days}$ and $\beta\rs{x}=0.8$; the last two parameters determine the fit of the \g-ray spectrum above $20\un{GeV}$. The emission spectrum for the same case but with the steady injection is shown by the green dashed line. The green dotted lines represent the spectra with the artificial injection terms which are the same as $Q\rs{t}$ but zero before the day 100 (the rightmost line), before the day 1517 (middle line) and before the day 8014 (the leftmost line). 
The \g-ray spectrum is calculated following the receipt of \citet{Aharonian-Atoyan-2000} with the cross-section for the pion production from \citet{Kelner-etal-2006}; such approach is  very accurate for \g-rays with energies below few $\un{TeV}$ \citep{Kelner-etal-2006}. 
               }
 \label{alpha:figfpp}
\end{figure}

Gamma-ray spectrum of IC443 (Fig.~\ref{alpha:figfpp}) is measured by the Fermi observatory \citep{Ackermann-etal-2013}, MAGIC \citep{Albert-etal-2007} and VERITAS \citep{Acciari-etal-2009}. 
The rapid decrease of the flux at energies below $\sim 200\un{MeV}$ discovered by \citet{Ackermann-etal-2013} is a sign of the hadronic \g-ray emission.

With our model, we confirm the hadronic origin of the \g-rays from IC443. Our fit is shown by the red solid line on Fig.~\ref{alpha:figfpp}. Though the data on $\alpha\rs{r}$ are fluctuating, the derived dependence $Q\rs{t}(t)$ and the \g-ray spectrum are rather smooth. 

Interpretation of the \g-ray spectrum of IC443 by \citet[][their Fig.~3]{Ackermann-etal-2013} has required the introduction of an ad-hoc, artificial spectral break in the proton spectrum around the proton energy $0.2\un{TeV}$. We obtain such a break (though at a somehow lower energy, $50\un{GeV}$ (it is around $p/m\rs{p}c\simeq 50$ on  Fig.~\ref{alpha:figsp}, red line) as a natural consequence of the time-dependent particle injection into the time-dependent acceleration. 

It merits to emphasize that the behavior of the term $Q\rs{t}$ is strictly determined by the radio observations only between the Days 1517-8014. Fig.~\ref{alpha:figfpp} clearly shows that this is exactly the period which is necessary to confirm that this break in the \g-ray spectrum is due to the injection variation: increasing photon energy corresponds to the change in $Q\rs{t}$ from more or less steady value after $\sim 10^4\un{day}$ to the approximate behavior $Q\rs{t}=\lg(t/950)$ during the period of the radio observations. The actual behavior of the radio index after the Day 8014 influences the part of the \g-ray spectrum below $2\un{GeV}$. 
Our fit follows the observed spectrum on the smaller photon energies. Therefore, our assumption about the steady-state injection after $\sim 10^4$ day seems to be right. 
From other side, our choice of the approximation $Q\rs{t}\propto t^{\beta\rs{x}}$ and the parameters $\beta\rs{x}$, $d\rs{x}$ which determine the behavior of $Q\rs{t}$ before Day $\approx 1500$ is less solid. They are chosen rather arbitrary, from the condition to trace the shape of the \g-ray spectrum above $\sim 20\un{GeV}$.

It is apparent Fig.~\ref{alpha:figfpp} that the steady injection (Fig.~\ref{alpha:figfpp}, green dashed line) cannot fit the observed \g-ray spectrum in IC443. 
The variation of the injection efficiency at the early times affects the energy distribution of \g-rays in SNR to a large extent. 

Which time periods are important in formation of certain portions of the \g-ray spectrum? The highest-energy protons require the longest time for acceleration. Therefore, the higher the photon energy, the earlier the times when the emitting particles were injected. It can be seen from Eqs.~(\ref{alpha:tm}) and (\ref{alpha:pm}) that the injection time and the momentum of particles at the present epoch are related as $t\rs{i}=(p/p\rs{m})^{-\alpha}t\rs{m}$. It is not so easy to relate the injection time to the energy of \g-rays. In order to answer the question, we calculated the \g-ray spectra also for the three artificial injection terms: they are the same as $Q\rs{t}(t)$ but zero before the day 100, the day 1517 and the day 8014 respectively. These spectra are shown by the green dotted lines on Fig.~\ref{alpha:figfpp}. In order to make the situation even more transparent, we mark by the vertical lines the photon energies where the contributions to the flux from the particles injected before these days become higher than $10\%$ (present time is about Day 11000). 

One can see for example that the change of the red line slope around the photon energy $0.3\un{TeV}$ (Fig.~\ref{alpha:figfpp}) is due to the switch from $Q\rs{t}\propto t^{\beta\rs{x}}$ to the steady injection at the day $d\rs{x}=100$. Thus, all the TeV hadronic gamma-rays from young and middle-age SNRs observed by the ground-based Cherenkov telescopes are from the protons injected mostly during {\it the first months} after SN explosion. The photon energy just below $20\un{GeV}$ corresponds to the time when the observations of the radio index were started. The energy about $2\un{GeV}$, near the point where all lines converge, corresponds to the last days of the radio data which we used. Thus, the \g-spectrum above $1\un{GeV}$ is determined by the injection during the first $20$ years of SN evolution. 

Clearly, the time dependence of injection at early times is very important in shaping the \g-ray spectra of SNRs.

\subsection{Can SN1987A be visible in the future in hadronic gamma-rays?}

Above, we have adopted SN1987A as a proxy for the IC443 supernova. Naturally, we may use $Q\rs{t}$ extracted from the evolution of the radio index in SN1987A to estimate also the visibility of SN1987A in hadronic \g-rays in the near future. 

The amplitude of the \g-ray spectrum is proportional to the product of the accelerated $n\rs{cr}$ and the target $n\rs{pp}$ proton densities $n\rs{cr}n\rs{pp}$. FERMI LAT did not detect SN1987A yet, therefore the density around SN1987A is smaller than the density around IC443. The age of SN1987A is $t\rs{sn87a}\simeq 0.01 t\rs{ic443}$. We consider two possible cases depending on the acceleration rate in SN 1987A.

\begin{figure}
 \centering
 \includegraphics[width=8truecm]{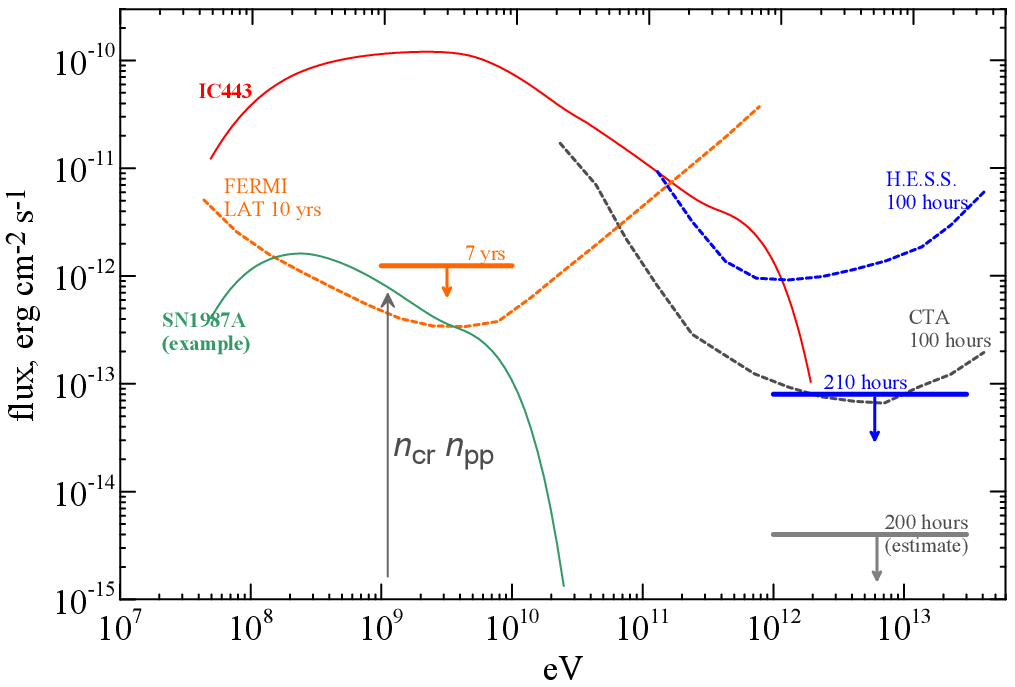}\\ 
 \includegraphics[width=8truecm]{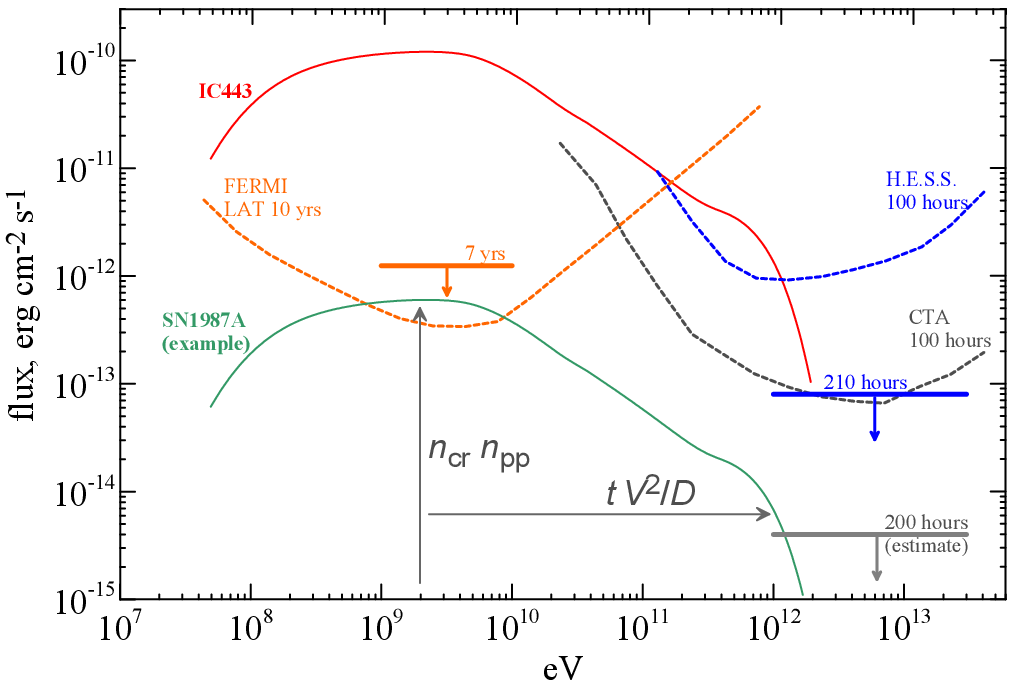}
 \caption{The hadronic \g-ray spectrum of SN1987A (green line). 
There are also shown: instrument sensitivities \citep{Funk-Hinton-2013}, the upper limits on the \g-ray flux after 7 years of FERMI observations \citep[orange,][]{Fermi-2016} and after 210 hours of H.E.S.S. observations \cite[blue,][]{HESS-2015}. The estimate of the upper limit for observations by CTA (gray) is obtained by scaling of the H.E.S.S. upper limit. The spectrum of IC443 (red line from Fig.~\ref{alpha:figfpp}) is also shown for comparison. 
The upper and lower plots corresponds to the \g-ray spectrum of SN1987A in the cases A and B described in the text.}
 \label{alpha:figappsn87a}
\end{figure}

\textit{Case A}. The acceleration rate in SN1987A is the same as in IC443 (Fig.~\ref{alpha:figappsn87a} upper plot). Let us consider an 'optimistic' value of the target protons in SN1987A $n\rs{pp,sn87a}\simeq 0.03 n\rs{pp,ic443}$ which could provide it to be detectable by FERMI in few years. The upper plot on Fig.~\ref{alpha:figappsn87a} demonstrate that, even if SN1987A will be detected in GeV \g-ray range, there is no chance to see the hadronic TeV \g-rays from SN1987A in the nearest future, also by CTA. This is because the available time (age of SN1987A) is not enough to accelerate protons to the energies which allow them to emit TeV \g-rays.  

\textit{Case B}. Acceleration in SN1987A is faster than in IC443 (Fig.~\ref{alpha:figappsn87a} lower plot). The right-hand portion of the \g-ray spectrum scales with time as $t/t\rs{acc}\propto tV^2/D$ where $t\rs{acc}$ is the acceleration time-scale. In order $t/t\rs{acc}$ to be larger in SN1987A, there are should be smaller diffusion coefficient and larger shock velocity, in order to provide 
\begin{equation}
\frac{t\rs{sn1987a}}{t\rs{ic443}}\frac{V\rs{sn1987a}^2}{V\rs{ic443}^2}\frac{D\rs{ic443}}{D\rs{sn1987a}}\geq 1.
\end{equation}
Thus, if FERMI detects SN1987A in few years and 
\begin{equation}
t\rs{acc,sn87a}\leq 0.01 t\rs{acc,ic443} 
\end{equation}
then SN1987A should be detectable in future by CTA (Fig.~\ref{alpha:figappsn87a} lower plot; here we used the 'optimistic' value of density $n\rs{pp,sn87a}\simeq 0.005 n\rs{pp,ic443}$).

\section{Conclusions}
\label{alpha:sect-concl}

We present an approach which could be a powerful tool to understand the evolution of SNe as well as the \g- and non-thermal X-rays from SNRs.

We outline the methodology and prove its validity on a clear test case, namely, the pair of SN1987A and IC443. The radio spectrum of the former is free of the self-absorption during the period of observations while the \g-spectrum of the later is hadronic in origin, i.e. it is not affected by the radiative losses as the leptonic \g-rays could.

We apply temporal evolution of the injection efficiency derived from the observed radio index of SN1987A to IC443. Is it correct to relate two different objects by the same function $Q\rs{t}$, especially because the progenitor of SN1987A is believed to be rather peculiar? We stress that $Q\rs{t}$, as it is shown in Sect.~\ref{alpha:sectradio}, depends on the average global trend in the evolution of the radio spectral index and is almost insensitive to the details which reflect the detailed configuration of the circumstellar medium around SN1987A. The main reason of the injection variation, as it is demonstrated in Sect.~\ref{alpha:reason}, is the increase of the surface of SN (i.e.  more particles cross the shock with a chance to be accelerated). Even more, the observed evolution of the radio index, as it is confirmed by the numerical model, corresponds to the period of time when the shock was in the nearly uniform H~\textsc{ii} region. We conclude therefore that the details of the configuration of the nebula around SN1987A are not effective in determination of the function $Q\rs{t}$. In addition, quite good correspondence between the radio data of SN1987A and the derived \g-ray spectrum of IC443, could be considered as a sign of similarities between the two remnants, of SN1987A and IC443. To conclude, since the main factor in the early evolution of the injection efficiency is the increase of the size of the very young remnant, the adoption of the $Q\rs{t}$ behavior extracted from SN1987A could therefore have general applicability, not strictly to SNR with exactly the same progenitor.

There are two basic results. First, the break in the proton spectrum in SNR IC443 around the proton energy $50\un{GeV}$ and the corresponding decrease of its \g-ray spectrum after $\sim 2\un{GeV}$ is, as proved by the radio observations of SN1987A, a consequence of the injection efficiency evolution during the supernova stage. 
Second, the shape of the \g-ray spectrum of IC443 at the photon energies above $20\un{GeV}$ results in a guess about a possible evolution of the injection in SN1987A before it became visible for the radio telescopes. Namely, the value $\beta\rs{x}=0.8$ which fits this part of the \g-ray spectrum of IC443 corresponds to the radio index $0.9$ during the unobserved period of SN1987A, which is quite reasonable index for supernovae.

We confirm the finding of \citet{Ackermann-etal-2013}, namely, that \g-rays from IC443 are of the hadronic origin. 
We found that the break in the proton spectrum around $50\un{GeV}$ (which is needed to explain the observed \g-ray spectra) naturally originates from the time-dependent injection during the early stage of the remnant evolution. Interesting that there are evidences for the same breaks in the proton spectra also in the SNR W44 \citep{Ackermann-etal-2013}, W51C \citep{Abdo-etal-2009,Jogler-Funk-2016} and W49B \citep{Abdalla-etal-2016}. These four SNRs also exhibit the specific hadronic low-energy cutoff in the \g-ray spectra (below $\sim 200\un{MeV}$) and are generally believed to be the remnants of the core-collapse SNe as well. 

There is a model for the origin of this proton spectrum break developed in the framework of the stationary particle acceleration \citep{Uchiyama-etal-2010,Cardillo-etal-2016}.  It does not explain, at the same time, the radio index variation on the supernova stage. Our scenario is in agreement with a general expectation that gamma-rays from SNRs like IC443 originate in the nearby molecular clouds. The clouds provide the target protons which generate \g-rays in interactions with the bullet protons which are accelerated by the SNR shock. In this framework, the \g-ray spectrum naturally reflects the momentum distribution of incoming relativistic protons (the ballets), not the particles in a cloud. The approach developed by \citet{Uchiyama-etal-2010,Cardillo-etal-2016} is completely alternative to ours: the shock re-accelerate the pre-existing cosmic rays. The early development of the particle acceleration during the SN stage is unimportant in this scenario.

The early time-dependent injection could explain as the shape of the \g-ray spectrum of SNR as the time evolution of the radio index in SN.
An alternative scenario for deviations of the synchrotron spectral indices in SNRs from the canonical value $0.5$ is the effect of the efficient non-linear particle acceleration; in this case, the shock compression factor $\sigma$ is modified. In particular, the non-stationary numerical model which accounts for the back-reaction of relativistic particles could also explain the evolution of the radio spectral index in SN1987A \citep[though with a lower accuracy, cf. Fig. 3 in ][ and our Fig.~\ref{alpha:figalpha}]{ber-ks-volk-2015} {\it together} with the variable injection \citep[dashed line on Fig.~3 in ][]{ber-ks-volk-2015}. The shock modification provides the second order corrections to the test-particle results. Our analytical description being test-particle in its nature catches the main trends and, due to its relative simplicity, clearly demonstrate that the observations of the radio spectral index evolution in SN1987A as well as the \g-ray spectrum of IC443 do not agree with the constant injection rate. Indeed, why should the injection be the same during different epochs and under different conditions? 
We stress that in the present paper we do not {\it model} the temporal evolution of the injection efficiency but {\it derive} it from observations of the radio index. Any model of the evolution of the radio emission should take the variable injection into account.

It worth to note once more that our solution of the time-dependent acceleration equation considers, for the first time, also the time-dependent injection (the original studies assume the steady-state injection, see \citet{Drury-1983} for the test-particle regime and \citet{Blasi-etal-2007} for the non-linear acceleration). The new effect in the present paper (change of the particle spectrum slope due to the variable injection) is just a consequence of the non-stationary particle injection into the non-stationary acceleration. The steady-state acceleration equation cannot catch the same effect even if the injection is time-dependent because the variable injection in this case affects the only normalization of the particle spectrum, Eq.~(\ref{kineq2:stationarysol}), through the factor $\eta$, -- but not its slope.

In summary, i) the time-scale for steady acceleration to $p\rs{max}$ is larger than an SNR age; ii) it is obligatory therefore to consider the non-stationary particle
spectrum for models of the $\gamma$-ray emission; iii) interpretation of the hard \g-ray spectra should take into account the evolution of the particle injection during the first months after the supernova event. And this is the point where SNe are linked to SNRs: information necessary to understand the \g-ray spectra of SNRs could be extracted the from present-day radio observations of SNe. The opposite is also true: actually, the \g-ray spectrum of SNR sheds light into its own first months after the supernova explosion.

\begin{acknowledgements}
We would like to thank G.Zanardo for the radio data on the spectral index evolution in SN1987A.  
This work is supported by the PRIN INAF 2014 grant `Filling the gap between supernova explosions and their remnants through magnetohydrodynamic modeling and high performance computing'. 
We acknowledge that the results in Sect.\ref{alpha:reason} have been achieved with the software developed in part by the U.S. Department of Energy supported Advanced Simulation and Computing/Alliance Center for Astrophysical Thermonuclear Flashes at the University of Chicago, using the PRACE Research Infrastructure resource MareNostrum III based in Spain at the Barcelona Supercomputing Center (PRACE Award N.2012060993). We acknowledge the CINECA Awards N.HP10BI36DG,2012 and HP10CKMKX1,2016 for the availability of high performance computing resources and support.
\end{acknowledgements}



\begin{appendix}
\section[]{Note on Equation~(8)}
\label{alpha:app1}

\begin{figure}[!h]
 \centering
 \includegraphics[width=8truecm]{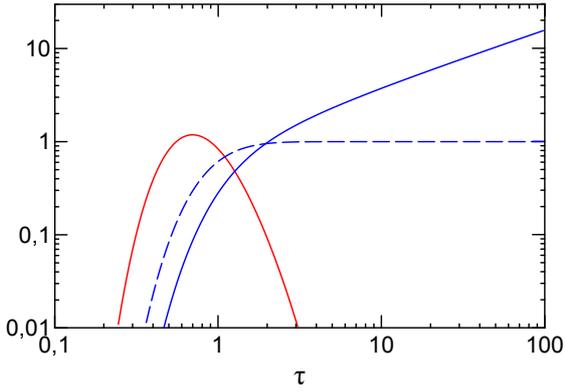}
 \caption{The same as Fig.~\ref{alpha:figprob} versus the dimensionless time $\tau$. The blue dashed line represents the integral in Eq.~(\ref{kineq2:solfTPQ}) with $Q\rs{t}=1$ and the solid blue line gives the same integral with $Q\rs{t}=\tau^{0.7}$.
               }
 \label{alpha:figapp}
\end{figure}

Let us consider the injection term in the form $Q\rs{t}\propto \tau^\beta$. The distribution function $\varphi_{\mathrm{o}}(\tau)$ (Fig.~\ref{alpha:figapp} red line) has prominent maximum around $\tau\rs{*}\approx 0.7\alpha^{-1}$ (the numerically evaluated values of $\tau\rs{*}$ are $0.70$ for $\alpha=1$, $1.6$ for $\alpha=1/2$ and $2.6$ for $\alpha=1/3$). 
After substitution Eq.~(\ref{kineq2:solfTPQ}) with $\varphi_{\mathrm{o}}(\tau')\approx \delta(\tau'-\tau\rs{*})$, we obtain, for $\tau>\tau\rs{*}$, that $f\rs{o}(t,p)\propto p^{-s\rs{f}}\left(\tau(t,p)-\tau\rs{*}\right)^{\beta}$. The spectral index of the particle spectrum is defined as
\begin{equation}
 s\equiv -\frac{d\ln f\rs{o}}{d\ln p}.
\end{equation}
Taking this derivative, with the use of the property $d\tau/\tau=-\alpha dp/p$, we have 
\begin{equation}
 s=s\rs{f}+\frac{\alpha\beta\tau}{\tau-\tau\rs{*}}.
 \label{alpha:app1eqs1}
\end{equation}
Radio emitting electrons in evolved SNe and SNRs have typically large $\tau$ (due to small $p$ and large $t$). For such $\tau\gg\tau\rs{*}$, Eq.~(\ref{alpha:app1eqs1}) transforms to Eq.~(\ref{alpha:sradiosnr}); this is demonstrated also by the blue solid line on Fig.~\ref{alpha:figapp}.
 
We consider the observations of SN1987A from $t=1517\un{day}$ and the scale $t\rs{m}=60\un{days}$. Therefore, the use of Eq.~(\ref{alpha:sradiosnr}) in the present paper is justified since we consider $\tau\geq 25$ and the value of $\tau\rs{*}=0.7$ makes negligible effect in Eq.~(\ref{alpha:app1eqs1}).

\end{appendix}

\end{document}